\documentclass[atoms,article,accept,moreauthors,pdftex,12pt,a4paper]{mdpi}
\usepackage{hyperref}
\usepackage{bbm}
\usepackage{amsmath, amssymb}
\usepackage{color}
\usepackage{graphicx,epsfig}
\usepackage{pifont}
\usepackage{amsmath, amssymb, graphics}

\lastpage{x}
\doinum{10.3390/------}
\pubvolume{xx}
\pubyear{2015}
\history{Received: xx / Accepted: xx / Published: xx}

\Title{Cavity assisted generation of sustainable macroscopic entanglement of ultracold gases}

\Author{Chaitanya Joshi$^{1}$* and Jonas Larson$^{2,3}$*}

\address{%
$^{1}$ Department of Physics and York Centre for Quantum Technologies, University of York, Heslington, York, YO10 5DD, UK\\
$^{2}$ Department of Physics, Stockholm University, Albanova physics center, Se-106 91 Stockholm, Sweden\\
$^{3}$ Institut f\"ur Theoretische Physik, Universit\"at zu K\"oln, De-50937, Köln, Germany}

\corres{chaitanya.joshi@york.ac.uk, jolarson@fysik.su.se.}

\abstract{Prospects for reaching persistent entanglement between two spatially separated atomic Bose-Einstein condensates  are outlined. The system set-up comprises of two condensates loaded in an optical lattice, which, in return, is confined within a high-$Q$ optical resonator. The system is driven by an external laser that illuminates the atoms such that photons can scatter into the cavity. In the superradiant phase a cavity field is established and we show that the emerging cavity mediated interactions  between the two condensates is capable of entangling them despite photon losses. This macroscopic atomic entanglement is sustained throughout the time-evolution apart from occasions of sudden deaths/births. Using an auxiliary photon mode and coupling it to a collective quadrature of the two condensates we demonstrate that the auxiliary mode's squeezing is proportional to the atomic entanglement and as such it can serve as a probe field of the macroscopic entanglement.}


\begin{document}


\section{Introduction}

After years of experimental activity in cavity QED mainly exploring fundamentals of light-matter interactions at a true quantum level~\cite{cqed}, in the second half of the 90's, these types of experiments took a new direction; it was demonstrated that they may constitute a platform for implementing quantum information processing~\cite{cavityqed}. In~\cite{cqed2}, for example, the light-matter interaction was monitored such that the photon state became entangled with a single two-level atom and then by letting a second atom `absorb' the light field a two-atom entangled state was obtained. In this protocol, after the first step quantum information is stored in the light field before it gets transmitted to the second atom. Such a scheme is inevitably sensitive to photon losses in the middle step, which, therefor, requires short operation times. It was, however, realized that the `cavity mediated atom-atom interaction' could also be utilized for directly entangling two atomic qubits without ever populating the lossy photon mode~\cite{cqed3}. In such a scenario, photons are only virtually excited, but nevertheless, integrating out the photon field results in an effective atom-atom interaction. Then, by carefully tuning the interaction time the desired entangling operation could be implemented. Somewhat surprisingly, even for a `classical' photon field, {\it i.e.} thermal state, it is indeed possible to use it for entanglement generation between the two atoms~\cite{cqed4}, and thus one may expect that the entanglement preparation is prone to both photon losses and other possible noise sources originating from the quantized radiation field.    

Following the pioneering cavity QED experiments, it has in recent years become possible to coherently couple not only a single atom but a gas of thousands of ultracold, condensed, bosonic atoms to a single quantized light mode~\cite{mbcqed1}. This scales up the effective light-matter interaction strength, and the strong coupling regime has indeed been demonstrated in several groups\footnote{For $N$ two-level atoms equally coupled to the field, the collective coupling goes as $\sqrt{N}$, and thus the characteristic time scale (inverse of the Rabi frequency) is greatly reduced.}~\cite{mbcqed2}. At these sub-Kelvin atomic temperatures, the atomic motion must be treated quantum mechanically -- it couples to the light field via the photon recoils. This led to explorations of a new, ultracold, regime of cavity QED. Much of the focus has especially been devoted to the collective phenomenon of {\it superradiance}\footnote{In this setting, superradiance does not necessarily manifest itself as an enhanced radiation process~\cite{mw} but instead as collective properties of the atomic condensate.}~\cite{mw}, realized by coupling the light degrees-of-freedom to the atomic vibrational modes rather than to internal atomic electronic states~\cite{mbcqed1}. Here, an atomic condensate is trapped inside the resonator and an external `perpendicular' drive/pump scatters photons into the cavity by illuminating the atoms. It turns out that at a critical pump amplitude, the atoms {\it self-organize} themselves in such a way that the different scattering processes constructively interfere making it possible for the photons to enter the resonator~\cite{domokos1}. Below the critical point, the condensate is approximately uniformly distributed and scattering processes add up destructively leading instead to a vanishing cavity field. This change in the system properties has shown to share many resemblances with the {\it Dicke phase transition} from a {\it normal} (vacuum cavity field) to a symmetry broken {\it superradiant phase} (coherent cavity light field)~\cite{dickept}. Properties of this transition has been the subject of several experimental activities\footnote{An alternative dynamical realization of the Dicke type transition is to consider internal electronic states Raman coupled via an external laser, first proposed in~\cite{carmichael} and later experimentally implemented in Ref.~\cite{dickeexp2}.}, for example the symmetry breaking has carefully been mapped out and also the critical exponents~\cite{dickeexp1,dickesym}. 

Like for the single/few atom cavity QED realizations, a natural question arises whether also these `many-body' cavity QED systems can serve as toolboxes for quantum information processing. A first step along these lines would be to study possibilities of entanglement generation among the many atoms. A crucial difference compared to the traditional cavity QED schemes is that these systems are driven. Thus, one should rather consider possible persistent entanglement in the long time limits of the driven-lossy system~\cite{ssent}. Just like for Refs.~\cite{cqed2,cqed3,cqed4}, the photon mode plays the role of an ancilla system  mediating atom-atom entanglement. The situation is different from those of~\cite{cqed3,cqed4} though since the non-vanishing field amplitude is sustained from balancing the gain from the input field to the losses due to photon decay, while in Refs.~\cite{cqed3,cqed4} the system is closed and the photon field initialized in some state. Furthermore, in the present setup the atomic fields will inevitably get entangled with the photon field, which causes an effective decoherence for the atomic state, {\it i.e.} the reduced state for the atoms will be mixed. In addition, photon dissipation gives rise to another decoherence channel for the atomic fields. All in all, even if in the few atom scenario quantum coherences of the radiation field is not necessary for atomic entanglement generation~\cite{cqed4}, due to the differences with the present many atomic situation it is {\it a priori} not clear whether any atom-atom entanglement may survive.

Inspired by a new system setup of the Esslinger group~\cite{eth} we analyze the above problem by considering two spatially separated atomic Bose-Einstein condensates (BECs) located within a cavity and perpendicularly driven by an external laser. Entanglement properties between the two atomic condensates (as previously mentioned the atomic degrees-of-freedom are collective atomic vibrational modes) are studied. As each condensate constitutes a macroscopic mechanical oscillator, the BEC-BEC entanglement is manifested in the corresponding vibrational states. By calculating the logarithmic negativity, persistent entanglement is demonstrated even in the open system case when photon losses are taken into account. More precisely, the BEC-BEC entanglement displays an infinite series of entanglement sudden deaths/births such that on average the two condensates keep their entanglement for an infinite time. Entanglement between the two condensates is, however, only possible in the superradiant regime. Close to and at the critical point, the main part of the quantum correlations are shared between the photon field and the atoms, while deeper in the superradiant phase  the BEC-BEC entanglement becomes more prominent. This is an example of entanglement sharing in a tripartite system~\cite{entshare}. In order to probe the entanglement between the two condensates we introduce an auxiliary photon mode and couple it to a collective quadrature of the two condensates. We demonstrate that near the critical point the squeezing of this auxiliary mode scales linearly with the the BEC-BEC entanglement and thus provides an indirect readout of the entanglement. However, losses of this additional photon mode implies that in the long time limit no quantum entanglement survives between the two condensates.


\section{Model system} 

By now, achieving a coherent light-matter coupling between a single BEC and a quantized photon mode has been demonstrated in several groups~\cite{mbcqed1,mbcqed2,dickeexp1,hemmerich,sk}. The coupling between single photons to the atomic collective vibrational modes realizes a longstanding goal of optomechanics, {\it i.e.} coherent interplay between photons and a macroscopic mechanical oscillator prepared in its ground state~\cite{om}. This has spur great activity in exploring hybrid quantum systems with the objective to devise scalable quantum architectures \cite{kham10}. In particular, generating nonclassical states in atom-photon coupled hybrid quantum systems has received significant theoretical and experimental interest \cite{htng09,dros14}. Continuing this quest, we envision two spatially separated BECs confined inside an optical resonator and explore whether macroscopic entanglement between two atomic BECs can be generated via the coupling to a common photon mode.

With the particular system in mind, the model Hamiltonian is given by~\cite{domokos1}
\begin{equation}\label{ham0}
\hat{H}=-\Delta\hat a^\dagger\hat a+\sum_{j=1,2}\int_0^L d{\bf x}\,\hat\Psi_j^\dagger({\bf x})\!\left[-\frac{\hbar^2}{2m}\nabla^2+U_0\hat a^\dagger\hat a\cos^2({\bf kx})+i\eta\cos({\bf kx})\left(\hat a^\dagger-\hat a\right)\right]\!\hat\Psi_j({\bf x}).
\end{equation}
Here, the parameter $\Delta$ is the cavity-pump detuning, $L$ is the cavity length, $m$ the atomic mass, ${\bf k}$ ($=(0,0,k)$) the wave number, $U_0=g_0^2/\Delta_a$ is the effective atom-cavity coupling with $g_0$ the bare Rabi frequency and $\Delta_a$ the atom-pump detuning, and $\eta=\Omega g_0/\Delta_a$ with $\Omega$ the pump amplitude. The photon creation (annihilation) operator is $\hat a^\dagger$ ($\hat a$), and $\hat\Psi_j({\bf x})$ is the atomic field operator annihilating an atom in condensate $j$ at position ${\bf x}$.  The operators obey $\left[\hat a,\hat a^\dagger\right]=1$, $\left[\hat\Psi_i({\bf x}'),\hat\Psi_j^\dagger({\bf x})\right]=\delta_{ij}$, and the remaining commutators are identically zero. In deriving (\ref{ham0}) the excited atomic electronic state has been adiabatically eliminated under the dispersive assumption $|\Delta|,\,|\Delta_a|\gg g_0$, and furthermore we consider a vanishing overlap between the two condensates such that the cross terms of the atomic operators can be neglected.

As a next step, we second quantize the Hamiltonian by expanding the field operators in the two lowest vibrational modes~\cite{domokos1,dickeexp1}
\begin{equation}\label{atf}
\hat\Psi_j({\bf x})=\frac{1}{\sqrt{L}}\hat c_0^{(j)}+\sqrt{\frac{2}{L}}\cos({\bf kx})\hat c_1^{(j)},
\end{equation}
where $\hat c_{0,1}^{(j)}$ annihilates an atom of vibrational mode $0,1$ in condensate $j$. Inserting the operators (\ref{atf}) into the expression (\ref{ham0}) for the Hamiltonian gives the low energy second quantized model, now containing five independent boson modes. Taking the conservation of number of atoms of the two condensates into account, it is practical to employ the Schwinger spin-boson mapping~\cite{schwing}
\begin{equation}
\begin{array}{lll}
\hat S_x=\displaystyle{\frac{1}{2}\left(\hat c_1^{(1)\dagger}\hat c_0^{(1)}+\hat c_0^{(1)\dagger}\hat c_1^{(1)}\right),} & \hspace{1cm} & \hat T_x=\displaystyle{\frac{1}{2}\left(\hat c_1^{(2)\dagger}\hat c_0^{(2)}+\hat c_0^{(2)\dagger}\hat c_1^{(2)}\right),}\\ \\
\hat S_y=\displaystyle{\frac{1}{2i}\left(\hat c_1^{(1)\dagger}\hat c_0^{(1)}-\hat c_0^{(1)\dagger}\hat c_1^{(1)}\right),} & \hspace{1cm} & \hat T_y=\displaystyle{\frac{1}{2i}\left(\hat c_1^{(2)\dagger}\hat c_0^{(2)}-\hat c_0^{(2)\dagger}\hat c_1^{(2)}\right),}\\ \\
\hat S_z=\displaystyle{\frac{1}{2}\left(\hat c_1^{(1)\dagger}\hat c_1^{(1)}-\hat c_0^{(1)\dagger}\hat c_0^{(1)}\right),} & \hspace{1cm} & \hat T_z=\displaystyle{\frac{1}{2}\left(\hat c_1^{(2)\dagger}\hat c_1^{(2)}-\hat c_0^{(2)\dagger}\hat c_0^{(2)}\right),}
\end{array}
\end{equation} 
to recast the Hamiltonian in a generalized Dicke model
\begin{equation}\label{hamltncavbecnonlin}
\hat{H}=-\Delta \hat{a}^{\dagger}\hat{a}+\omega_{R}(\hat{S}_{z}+\hat{T}_{z})
+\frac{g}{\sqrt{N}}(\hat{a}^{\dagger}+\hat{a})(\hat{S}_{+}+\hat{S}_{-})+\frac{g}{\sqrt{N}}(\hat{a}^{\dagger}+\hat{a})(\hat{T}_{+}+\hat{T}_{-})+ \frac{U}{N}\hat{a}^{\dagger}\hat{a}(\hat{S}_{z}+\hat{T}_{z}),
\end{equation}
where $\omega_R=\hbar k^2/2m$ is the recoil frequency. The atomic operators $\{\hat{S}_{\pm},\hat{S}_{z},\hat{T}_{\pm},\hat{T}_{z}\}$ satisfy the $SU(2)$ commutation relations $[\hat{J}_{\pm},\hat{J}_{z}]=\mp \hat{J}_{\pm}, [\hat{J}_{+},\hat{J}_{-}]=2\hat{J}_{z}$, where $\hat{J} \in \{\hat{S},\hat{T}\}$. The standard Dicke model is lacking the last term and the boson mode couples only to a single spin $S$\footnote{Since the spins are preserved we could define a collective spin $R=S+T$ and thereby recover the regular Dicke spin-boson interaction.}.  We have introduced the total number of atoms $N$; the effective atom-light coupling $g=\sqrt{2N}\eta$ and the Stark shift term $U=NU_0/4$. With this parameter scaling, for large atom numbers every term in the Hamiltonian $\mathcal{O}(N)$. This ensures that in the thermodynamic limit, $N,\,L\rightarrow\infty$ and $N/L=$constant, the model exhibits a second order phase transition~\cite{dickept}. In the following we will omit the Stark shift term, {\it i.e.} setting $U=0$, valid in the regime were the pump amplitude is much larger than the Rabi frequency which is justified in typical experiments~\cite{dickeexp1}. In absence of any losses the coupling $\tilde{g_c}=\sqrt{|\Delta|\omega_R}/2$ separates the two phases, normal ($g<\tilde{g_c}$) from the superradiant ($g>\tilde{g_c}$) one. In the normal phase, the ground state is simply the vacuum state of all modes ($\hat S_z=\hat T_z=-N/4$). In the superradiant phase, however, a $\mathbb{Z}_2$ parity symmetry is spontaneously broken which is for instance reflected in a 0 or $\pi$ phase of the photon field~\cite{dickesym}. The spins in this phase are tilted away from the south pole toward the `$x$'-direction~\cite{jonas}. Classically, the phase transition manifest as a pitchfork bifurcation~\cite{milburn2}, and it has been argued that entanglement generation should be most pronounced in the symmetry broken phase~\cite{milburn}. The cavity induced atom-atom interaction which underpin the entanglement is transparent when the cavity field is eliminated giving rise to an effective atomic model of the Lipkin-Meshkov-Glick type~\cite{jonas2}. Entanglement properties of such models have been studied in the Refs.~\cite{jonas2,dros14}. Integrating out the photon field and imposing an adiabatic approximation in order to derive a Lipkin-Meshkov-Glick model is not always justified, for example for moderate detunings $\Delta$ and good cavities (low photon decay rates $\kappa$). Clearly, in this case the full system has to be considered. This may be extra important when studying entanglement properties since the presence of the cavity field will itself get entangled with the atomic fields and thereby indirectly degrade atom-atom entanglement. In addition, the photons may leak out of the cavity which result in an additional decoherence mechanism for the atoms.

To take photon decay into account we follow the standard procedure by coupling the cavity mode to a set of harmonic bath oscillators and apply the Born-Markov and secular approximations that are supposed to be valid in the optical regime~\cite{hpbr02}. Doing so we end up with a Markovian master equation of the Lindblad form. Assuming a zero temperature bath the resulting open dynamics can be modeled through the following master equation 
\begin{eqnarray}\label{msterqnneq}
\frac{d}{dt}\hat\rho=-i[\hat H,\hat\rho]+\kappa\hat{\mathcal L}_{\hat{a}}[\hat\rho],
\end{eqnarray}
where $\hat{\mathcal L}_{\hat{x}}[\hat \rho]=2 \hat x \rho \hat x^\dagger - \hat x^\dagger \hat x \rho - \rho \hat x^\dagger \hat x$ is a Lindblad superoperator~\cite{hpbr02} and $\hat\rho$ is the full system density operator. We should keep in mind that we consider a driven system out of equilibrium. This justifies a phenomenological introduction of losses as modeled through the master equation \eqref{msterqnneq}.  In particular, the light-matter coupled  Hamiltonian \eqref{hamltncavbecnonlin} is written in the frame of an external drive imparting  it a non-equilibrium character. For a time-independent system coupled to an external bath it is required that the steady state will obey the principles of equilibrium statistical mechanics, while here no such restriction is enforced on the dynamics which arises out of an implicit time-dependent Hamiltonian~\cite{cjos14}.

\subsection{Semiclassical analysis} 

As we are interested in entanglement generation on a macroscopic level we assume the number of atoms $N\gg1$. This is typically the case in experimental realizations were $N\gtrsim10^5$~\cite{dickeexp1,hemmerich,dickesym,roton}. For particle numbers of this size, mean-field approaches qualitatively captures many of the system characteristics~\cite{dickeexp1}. However, such a semiclassical mean-field approach is incapable of capturing or quantifying the quantum correlations. For bosonic (spin) systems, such quantum corrections are conveniently explored in terms of the Holstein-Primakoff representation~\cite{hp}, from which we can extract information about the entanglement. Before analyzing the quantum features we first need to determine the mean-field solutions of the model which are the steady state solutions of the semiclassical equations of motion governed by the master equation~\eqref{msterqnneq}.

Let us define  $\langle \hat{a} \rangle=\alpha$, $\langle \hat{S}_{-} \rangle=\beta, \langle \hat{T}_{-} \rangle=\delta, \langle \hat{S}_{z} \rangle=w_{\hat{S}}$, and $\langle \hat{T}_{z} \rangle=w_{\hat{T}}$. The semiclassical equations of motion are given from the Heisenberg equations under the factorization assumption, i.e., $\langle (\hat{a}^{\dagger}+\hat{a})(\hat{J}_{-}-\hat{J}_{+})\rangle \rightarrow \langle (\hat{a}^{\dagger}+\hat{a})\rangle \langle(\hat{J}_{-}-\hat{J}_{+})\rangle, \langle (\hat{a}^{\dagger}+\hat{a})\hat{J}_{z}\rangle \rightarrow \langle (\hat{a}^{\dagger}+\hat{a})\rangle \langle \hat{J}_{z}\rangle$. This results in a finite set of coupled nonlinear equations
\begin{eqnarray}
\dot{\alpha}&=&-(\kappa-i\Delta)\alpha-i\frac{g}{\sqrt{N}}(\beta+\beta^{*}+\delta+\delta^{*}), \\
\dot{\beta}&=&-i\omega_{R}\beta+2i\frac{g}{\sqrt{N}}(\alpha+\alpha^{*})w_{\hat{S}},\\
\dot{w}_{\hat{S}}&=&i\frac{g}{\sqrt{N}}(\alpha+\alpha^{*})(\beta-\beta^{*}),\\
\dot{\delta}&=&-i\omega_{R}\delta+2i\frac{g}{\sqrt{N}}(\alpha+\alpha^{*})w_{\hat{T}},\\
\dot{w}_{\hat{T}}&=&i\frac{g}{\sqrt{N}}(\alpha+\alpha^{*})(\delta-\delta^{*}).
\end{eqnarray}
At this level we do not include any back-action of the reservoir apart from the dissipative photon decay. This is justified when the cavity mode is coupled to a heat-bath in thermal equilibrium and the associated  Langevin noise operators thus satisfy $\langle \hat{a}_{\rm in}(t) \rangle =0$~\cite{carm}. Spin conservation for each atomic BEC at the mean-field level takes the form 
\begin{eqnarray}
w_{\hat{T}}^{2}+|\beta|^{2}&=&\frac{N^{2}}{4},\\
w_{\hat{S}}^{2}+|\delta|^{2}&=&\frac{N^{2}}{4}.
\end{eqnarray} 
Making use of the above equations we can find the steady state solutions for $\alpha,\beta,\delta,w_{\hat{T}},w_{\hat{S}}$. We find that the solution predicts the occurrence of two distinct  phases (normal and superradiant) which are separated by a critical value of the light-matter coupling $g_{c}=\frac{\sqrt{\omega } \sqrt{\Delta ^2+\kappa ^2}}{2 \sqrt{2}\sqrt{|\Delta| }}$.
 Note how the critical coupling is modified by the photon decay rate $\kappa$~\cite{domokos1,carmichael,keeling}. The open nature of the system not only shifts the transition, but also alters the critical exponents~\cite{open}.

For $g<g_{c}$, the steady state solution is a fully inverted (trivial) state with $\alpha=\beta=\delta=0, w_{\hat{T}}=w_{\hat{S}}=-N/2$. This represents the normal phase. For $g>g_{c}$, however, this trivial state becomes unstable and two new stable steady states arise with
   \begin{eqnarray}\label{meanf}
   \begin{aligned}
   \alpha=\mp2i\frac{g\sqrt{N}}{\kappa-i\Delta}\sqrt{1-\left(\frac{g_{c}}{g}\right)^{4}},\\
   \beta=\delta=\pm\frac{N}{2}\sqrt{1-\left(\frac{g_{c}}{g}\right)^{4}},\\
    w_{\hat{T}}= w_{\hat{S}}=-\frac{N}{2}\left(\frac{g_{c}}{g}\right)^{2}.
    \end{aligned}
   \end{eqnarray}
This is the aforementioned pitchfork bifurcation~\cite{milburn2}. In particular, we will show later that a consequence of the light-matter coupling exceeding a critical value ($g/g_{c}>1$) is the generation of bi-partite entanglement between the two distant BEC samples, while they remain separable in the normal phase.

\subsection{Linearized light-matter interaction}

As argued above, in the large particle regime the quantum fluctuations are small  and may be treated in a linearized approach. Starting with the normal phase we use the Holstein-Primakoff representation~\cite{hp}
\begin{eqnarray}
\hat{S}_{-}&=&\sqrt{N-\hat{b}^{\dagger}\hat{b}}\hat{b},\\
\hat{S}_{z}&=&\hat{b}^{\dagger}\hat{b}-\frac{N}{2},\\
\hat{T}_{-}&=&\sqrt{N-\hat{c}^{\dagger}\hat{c}}\hat{c},\\
\hat{T}_{z}&=&\hat{c}^{\dagger}\hat{c}-\frac{N}{2},
\end{eqnarray}
where $\hat b$ and $\hat c$ are the two new boson modes describing the bosonic excitations around the mean-fields (\ref{meanf}), to reformulate the Hamiltonian~\eqref{hamltncavbecnonlin} in terms of three boson fields. By further making use of the large $N$ limit, we linearize the resulting Hamiltonian to arrive at
\begin{equation}\label{hamtfrnorml}
\hat H_{\rm n}^{\rm lin}=-\Delta \hat{a}^{\dagger}\hat{a}+\omega_{R}(\hat{b}^{\dagger}\hat{b}+\hat{c}^{\dagger}\hat{c})+g(\hat{a}^{\dagger}+\hat{a})(\hat{b}^{\dagger}+\hat{b}+\hat{c}^{\dagger}+\hat{c}).
\end{equation}
In the superradiant  phase ($g/g_{c}>1$) the cavity and the atomic field modes are macroscopically excited and the expansion of the Hamiltonian \eqref{hamltncavbecnonlin} is preceded by first displacing the operators $\hat{a}, \hat{b}, \hat{c}$ around their steady state amplitudes (\ref{meanf}). This gives the following linearized Hamiltonian
\begin{eqnarray}\label{hamtfrsuper}
\hat H_{\rm sr}^{\rm lin}=-\Delta \hat{a}^{\dagger}\hat{a}+\Omega(\hat{b}^{\dagger}\hat{b}+\hat{c}^{\dagger}\hat{c})+\zeta(\hat{b}^{2}+\hat{b}^{\dagger 2}+\hat{c}^{2}+\hat{c}^{\dagger 2}) +\phi (\hat{a}^{\dagger}+\hat{a})(\hat{b}^{\dagger}+\hat{b}+\hat{c}^{\dagger}+\hat{c}),
\end{eqnarray}
where 
\begin{eqnarray}
\Omega&=&\frac{\omega_{R}}{2 \mu}(1+\mu)+\frac{\omega_{R}(1-\mu)(3+\mu)}{4 \mu(1+\mu)},\nonumber \\
\zeta&=&\frac{\omega_{R}(1-\mu)(3+\mu)}{8 \mu(1+\mu)},\nonumber \\
\phi&=&g\mu \sqrt{\frac{2}{1+\mu}}\nonumber, \\
\mu&=&{\left(\frac{g_{c}}{g}\right)}^{2}.\nonumber 
\end{eqnarray}
Note that in the limit $g\rightarrow g_c$, $\hat H_{\rm sr}^{\rm lin}$ equals $\hat H_{\rm n}^{\rm lin}$. In these linearized forms, both the Hamiltonians are quadratic for which entanglement properties have been thoroughly studied~\cite{vedral}. For example, the logarithmic negativity $\rm N$~\cite{gera07} used in the next section is a good measure in order to quantify the bi-partite entanglement. To further simplify the Hamiltonians \eqref{hamtfrnorml} and \eqref{hamtfrsuper} we define the following set of collective operators
\begin{eqnarray}\label{colbas}
\hat{p}&=&(\hat{b}+\hat{c})/2+\hat{a}/\sqrt{2},\nonumber \\
\hat{q}&=&(\hat{b}+\hat{c})/2-\hat{a}/\sqrt{2},\nonumber \\
\hat{s}&=&(\hat{b}-\hat{c})/\sqrt{2}\nonumber, 
\end{eqnarray}
or, inversely 
\begin{eqnarray}
\hat{a}&=&(\hat{p}-\hat{q})/\sqrt{2},\nonumber \\
\hat{b}&=&(\hat{p}+\hat{q})/2+\hat{s}/\sqrt{2},\nonumber \\
\hat{c}&=&(\hat{p}+\hat{q})/2-\hat{s}/\sqrt{2}\nonumber.
\end{eqnarray}
In terms of these the Hamiltonian in the normal phase \eqref{hamtfrnorml} takes the form 
\begin{eqnarray}\label{clctvhamtfrnorml}
\hat{H}_{\rm n}^{\rm lin} = \tilde{\omega_{p}}\hat{p}^{\dagger}\hat{p}+\tilde{\omega_{q}}\hat{q}^{\dagger}\hat{q}+\omega_{R}\hat{s}^{\dagger}\hat{s}+\tilde{g}(\hat{p}^{\dagger}\hat{q}+\hat{q}^{\dagger}\hat{p}) +\frac{g}{\sqrt{2}}(\hat{p}^{2}+\hat{p}^{\dagger 2}-\hat{q}^{2}-\hat{q}^{\dagger 2}),
\end{eqnarray}
with 
\begin{eqnarray}
\tilde{\omega_{p}}&=&\frac{\omega_{R}-\Delta}{2}+\sqrt{2}g,\nonumber \\
\tilde{\omega_{q}}&=&\frac{\omega_{R}-\Delta}{2}-\sqrt{2}g,\nonumber \\
\tilde{g}&=&\frac{\omega_{R}+\Delta}{2}.
\end{eqnarray}
Similarly, the Hamiltonian in the superradiant  phase \eqref{hamtfrsuper} can be expressed as 
\begin{eqnarray}\label{clctvhamtfrsuper}
\tilde{H}_{\rm sr}^{\rm lin} &=&\omega_{p}\hat{p}^{\dagger}\hat{p}+\omega_{q}\hat{q}^{\dagger}\hat{q}+\Omega\hat{s}^{\dagger}\hat{s}+\frac{(\Omega+\Delta)}{2}(\hat{p}^{\dagger}\hat{q}+\hat{q}^{\dagger}\hat{p})
+\zeta (\hat{p}\hat{q}+\hat{q}^{\dagger}\hat{p}^{\dagger})+\zeta(\hat{s}^{2}+\hat{s}^{\dagger 2})\nonumber \\
&&+\left(\frac{\zeta}{2}+\frac{\phi}{\sqrt{2}}\right)(\hat{p}^{2}+\hat{p}^{\dagger 2})+\left(\frac{\zeta}{2}-\frac{\phi}{\sqrt{2}}\right)(\hat{q}^{2}+\hat{q}^{\dagger 2}),
\end{eqnarray}
where 
\begin{eqnarray}
\omega_{p}&=&\frac{\Omega-\Delta}{2}+\sqrt{2}\phi,\nonumber \\
\omega_{q}&=&\frac{\Omega-\Delta}{2}-\sqrt{2}\phi.\nonumber 
\end{eqnarray}
From the above expressions it is seen that the relative mode $\hat s$ decouples from the remaining two modes. 

So far, the open character of the system only enters in the system parameters of the above effective low energy models. We have not taken the Lindblad terms into account. Since losses only appear for the photon mode, the Lindblad term of Eq. (\ref{msterqnneq}) is unaltered going to the Holstein-Primakoff representation. Nevertheless, working in the collective basis (\ref{colbas}) we should also transform the loss term; 
\begin{eqnarray}\label{collctvmeq}
\frac{d}{dt}\hat\rho&=&-i[\hat{H},\hat\rho]+\frac{\kappa}{2}\hat{\mathcal L}_{\hat{p}}[\hat \rho]+\frac{\kappa}{2}\hat{\mathcal L}_{\hat{q}}[\hat\rho]
-\frac{\kappa}{2}(2 \hat p \hat\rho \hat q^\dagger - \hat q^\dagger \hat p \hat\rho - \hat\rho \hat q^\dagger \hat p)
-\frac{\kappa}{2}(2 \hat q \hat\rho \hat p^\dagger - \hat p^\dagger \hat q \hat\rho - \hat\rho \hat p^\dagger \hat q),
\end{eqnarray}
where $\hat H=\hat{H}_{\rm n}^{\rm lin}$ or $\hat H=\hat{H}_{\rm sr}^{\rm lin}$ depending on the phase studied. Note that since the $\hat s$ mode is purely atomic it does not couple to the other modes even when the system interacts with the surrounding reservoir. In particular, in the normal phase of the closed system the ground state for the $\hat s$ mode is the vacuum. In the superradiant phase, however, the corresponding ground state is a squeezed vacuum~\cite{mw}. Thus, if the system's $\hat s$ mode is initialized in vacuum it is not an eigenstate and in particular it evolves accordingly
\begin{equation}\label{coreq}
\begin{array}{l}
\displaystyle{\langle \hat{s}^{\dagger}\hat{s} \rangle = -\frac{4 \zeta ^2 \sin ^2\left(t \sqrt{\Omega ^2-4 \zeta ^2}\right)}{4 \zeta ^2-\Omega ^2},} \\ \\
\displaystyle{\langle \hat{s}^{\dagger 2} \rangle=\frac{\zeta  \left(\Omega ^2-4 \zeta ^2\right) \sin \left(2t \sqrt{\Omega ^2-4 \zeta ^2}\right)}{\left(4 \zeta ^2-\Omega ^2\right)^{3/2}}+\frac{2 i \zeta  \Omega  \sqrt{\Omega ^2-4 \zeta ^2} \sin ^2\left(t \sqrt{\Omega ^2-4 \zeta
   ^2}\right)}{\left(4 \zeta ^2-\Omega ^2\right)^{3/2}}.}
\end{array}
\end{equation}
A consequence of this time-dependence of the $\hat s$ mode is that correlators involving the  modes $\hat b$ and $\hat c$ may also become time-dependent, even though the collective modes $\hat p$ and $\hat q$ may approach a stable fixed point. For instance, if the $\hat a$, $\hat b$, and $\hat c$ modes are initialized in their respective vacuum states then in the 
steady state one gets 
\begin{eqnarray}\label{eqnfrmdesbc}
\langle \hat{b}^{\dagger}\hat{b} \rangle_{\rm ss}=\langle \hat{c}^{\dagger}\hat{c} \rangle_{\rm ss}=\frac{1}{4}\langle \hat{p}^{\dagger}\hat{p}+\hat{q}^{\dagger}\hat{q}+\hat{p}^{\dagger}\hat{q}+\hat{q}^{\dagger}\hat{p} \rangle_{\rm ss}-\frac{1}{2}\frac{4 \zeta ^2 \sin ^2\left(t \sqrt{\Omega ^2-4 \zeta ^2}\right)}{4 \zeta ^2-\Omega ^2}.
\end{eqnarray}
Similar expressions can be obtained  for other correlators containing the $\hat{b}$ and $\hat{c}$ operators. 

\section{Macroscopic entanglement}
\subsection{Sustainable entanglement generation}

It should be remarked that the master equations \eqref{collctvmeq}  preserve the initial Gaussian character of the modes $\hat p$, $\hat q$, and $\hat s$ (or, equivalently of modes $\hat a$, $\hat b$, and $\hat c$). Using standard quantum optical techniques it allows us to convert  \eqref{collctvmeq} into  partial differential equations for the quantum characteristic function defined as $\chi(\epsilon_{p},\epsilon_{q})=\langle e^{\epsilon_{p}\hat{p}^{\dagger}}e^{-\epsilon_{p}^{*}\hat{p}}e^{\epsilon_{q}\hat{q}^{\dagger}}e^{-\epsilon_{q}^{*}\hat{q}}\rangle$. We will not enter into the details of this exercise, but will refer the reader to the same approach described  in \cite{cjos14}. As pointed out earlier, the solutions verify that the $\hat b$ and $\hat c$ modes remain separable throughout the normal phase. In the superradiant phase, however, the atomic $\hat b$ and $\hat c$ modes can become entangled\footnote{We also find that the $\hat a$ and $\hat b$ modes (or, equivalently, modes $\hat a$ and $\hat c$) do exhibit bi-partite entanglement both in the normal and superradiant phases.}. We characterize the bi-partite  entanglement between the various modes in terms of the logarithmic negativity $\rm N$\cite{gera07}.

\begin{figure}[h!]
  \centering
    \includegraphics[width=0.7\textwidth]{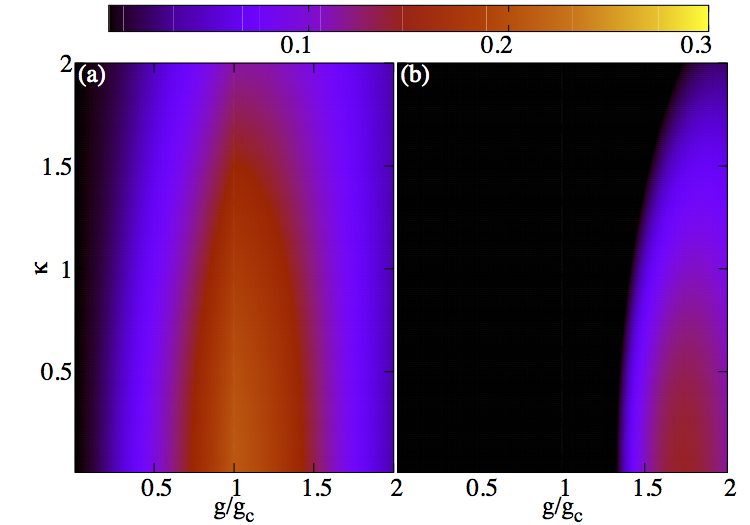}\\
     \includegraphics[width=0.7\textwidth]{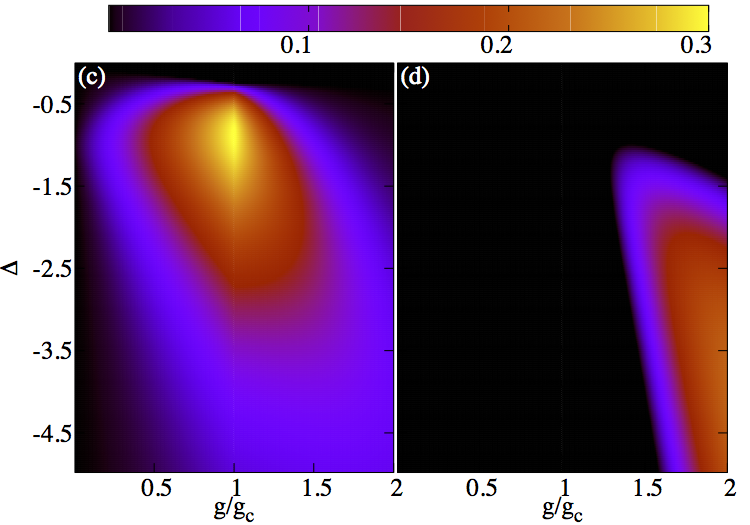}
   \caption{Bi-partite entanglement (logarithmic negativity $\rm N$) between  the cavity mode and each atomic BEC, (a) and (c), and bi-partite entanglement between the two  atomic  BECs, (b) and (d). The fact that the cavity-BEC or the BEC-BEC entanglement roughly complement one another in terms of amplitude is an outcome of entanglement sharing~\cite{entshare}. We also note that atom-atom entanglement occurs only in the superradiant phase and 
beyond a critical value of the detuning, {\it i.e.} $|\Delta|>\Delta_{c}$, {\it i.e.} it is not symmetric with respect to the photon-atom detuning $\tilde{\delta}=-\Delta-\omega_R$. In (a) and (b) $\Delta=-2$, while in (c) and (d) $\kappa=0.05$ and for all four plots $t=k\pi/\sqrt{\Omega ^2-4 \zeta ^2}$, $k \in \mathbb{Z}$. The value of the loss rate $\kappa$ has been taken small, but still experimentally relevant~\cite{dickeexp1,dickesym,roton}, in order to achieve as large entanglement as possible. The (scaled) recoil frequency $\omega_{R}=1$.   }
 \label{ent_modesbcvskappadelta}
\end{figure}

For times $t =k\pi/\sqrt{\Omega ^2-4 \zeta ^2}$, $k \in \mathbb{Z}$  the $\hat{b}$ and $\hat{c}$ modes do not directly couple to the $\hat s$ mode.  Bi-partite entanglement between the $\hat{b}$ and $\hat{c}$ modes for such a choice of $t$ is shown in Fig.~\ref{ent_modesbcvskappadelta} (b) and (d). Also shown  in Fig.~\ref{ent_modesbcvskappadelta} (a) and (c) is the bi-partite entanglement between the cavity mode and one of the atomic BECs. As can be seen from Fig.~\ref{ent_modesbcvskappadelta} (a) and (c), bi-partite entanglement develops between the cavity field and each atomic cloud. This is true in the normal and superradiant phase. However, it is only when $g/g_{c}>1$ that the entanglement between distant atomic BECs develops. Entanglement between the cavity field and each atomic cloud is maximal near the critical point $g/g_{c}=1$, which may serve as an indicator of the quantum critical point. This is indeed a universal behavior; at the critical point where the length scales (in system with short range interactions) diverge the entanglement also diverges~\cite{entcrit}. Furthermore, as a function of the effective detuning $\tilde{\delta}=-\Delta-\omega_R$ the maximum entanglement is established at the critical point and at resonance $\tilde{\delta}=0$. It is interesting to observe that in our model, which does not belong to the class of short range interaction ones, the bi-partite entanglement reaches its maximum at the critical points only in the cavity-BEC subsystems, but not in the BEC-BEC subsystems. In fact, as bi-partite BEC-BEC entanglement builds up (increasing the coupling strength) the cavity-BEC entanglement decreases. This is the phenomenon of entanglement sharing~\cite{entshare}, which for a tripartite system implies that simultaneous large bi-partite entanglement among all constituents is not allowed. We note that for a special case of our model this effect has been demonstrated namely for the closed Tavis-Cummings model ({\it i.e.} the Dicke model with the rotating wave approximation) for two two-level atoms~\cite{entshare2}. An additional feature evident from Fig.~\ref{ent_modesbcvskappadelta} is that BEC-BEC entanglement appears only beyond a critical value of the detuning, {\it i.e.} $\Delta<\Delta_{c}$. In other words, being in the superradiant regime is not always sufficient to warrant BEC-BEC entanglement. 

\begin{figure}[h!]
  \centering
     \includegraphics[width=0.4\textwidth]{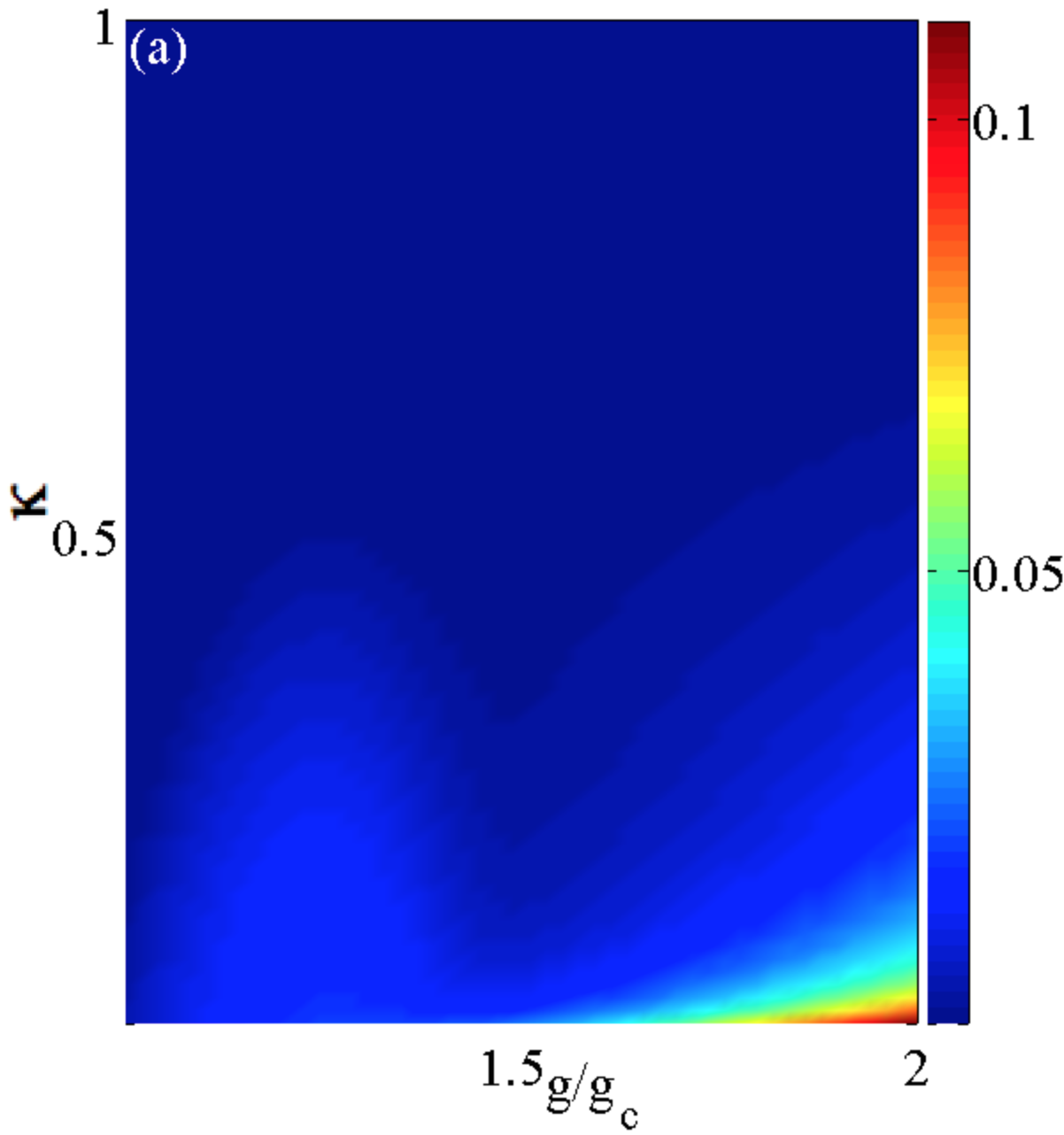}
     \includegraphics[width=0.4\textwidth]{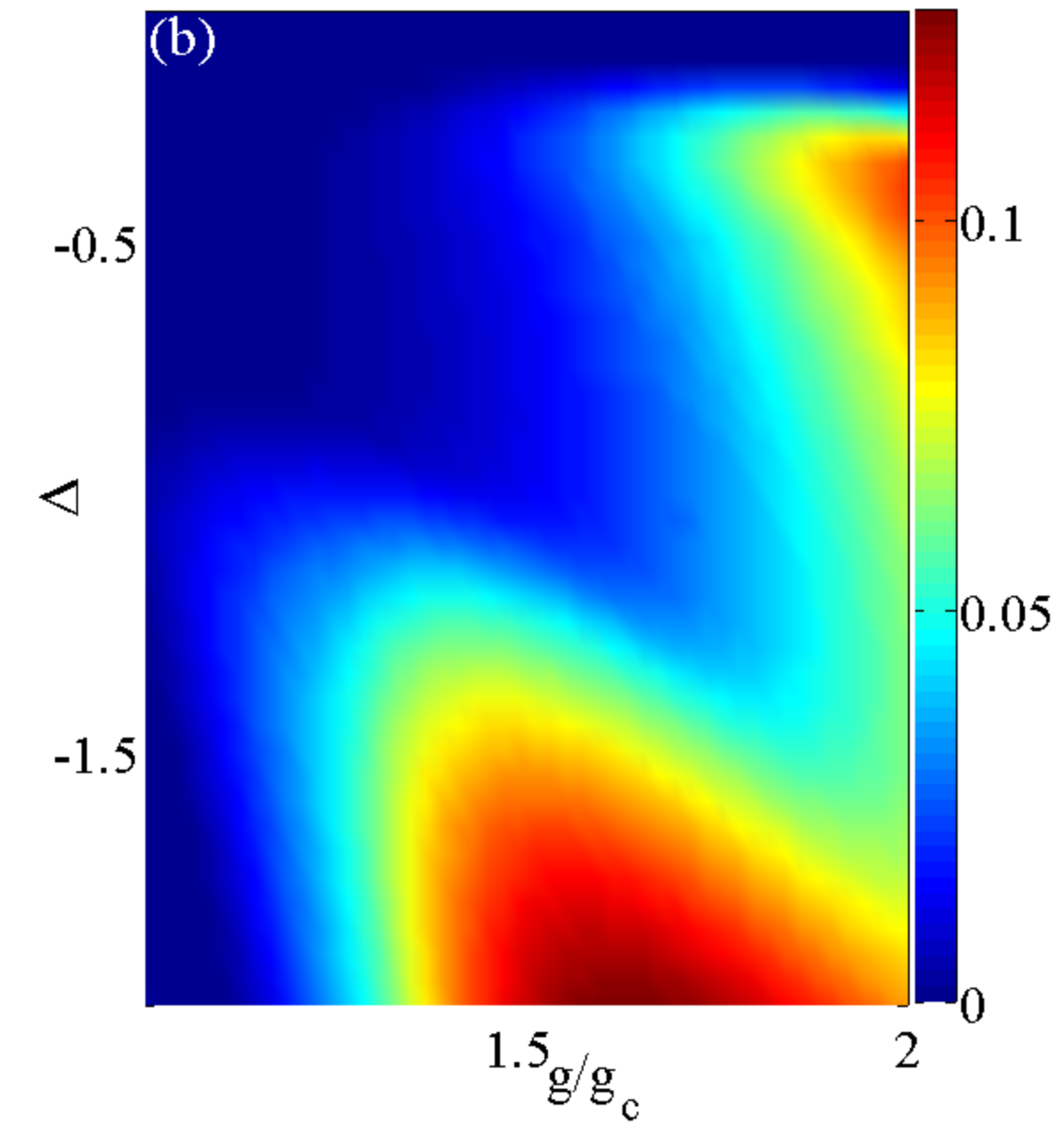}\\
      \includegraphics[width=0.8\textwidth]{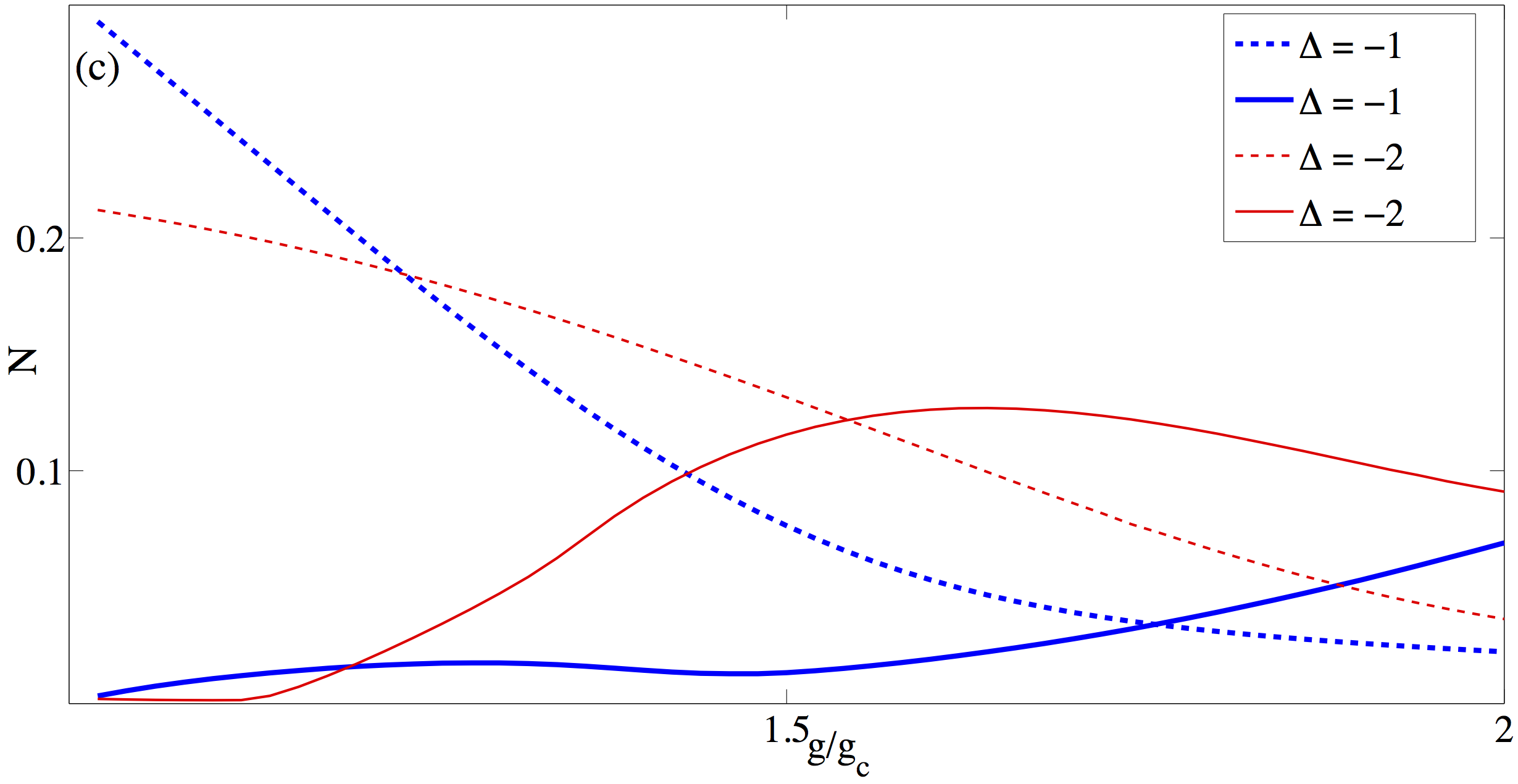}
   \caption{
Time-averaged bi-partite entanglement (logarithmic negativity $\rm N$) between the two condensates as a function of $(a)$ $\kappa$ and $g/g_{c}$ for $\Delta$ = -1, $(b)$ $\Delta$ and $g/g_{c}$ for $\kappa$ = 0.05. Shown in $(c)$ is the time-averaged cavity-BEC (dashed) and BEC-BEC (solid) entanglement for two different values of the parameter $\Delta$ and $\kappa$ = 0.05. The rise of atomic entanglement as the atom-photon entanglement decreases is again an outcome of the entanglement sharing mechanism. Other physical parameters include $\omega_{R}=1$.}
 \label{time_avg_no_aux}
\end{figure}

The results reported in Fig.~\ref{ent_modesbcvskappadelta} are obtained for a particular choice of parameter $t$. Experimentally it is more relevant to ask the question how the time-averaged entanglement depends on the system parameters. One may, for example, imagine a continuous measurement and then averaging the results, or making measurements at random times and then averaging. Thus, we evaluate the time-averaged logarithmic negativity as a measure of bi-partite entanglement among photon and atomic modes, {\it i.e.} we evaluate ${\rm N}=\int_{0}^{T}{\rm N}(t) dt/T$ for large sampling times $T$'s. The resulting time-averaged logarithmic negativity is plotted  in Fig.~\ref{time_avg_no_aux}. As is evident from Fig.~\ref{time_avg_no_aux} (a), the time-averaged bi-partite entanglement between the two BEC samples is indeed very fragile to cavity damping. As illustrated  in Fig.~\ref{time_avg_no_aux} (b), the  time-averaged bi-partite entanglement between the two BECs is a nontrivial function of the cavity field detuning $\Delta$ and the coupling strength $g/g_{c}$. In particular, the two BECs may become entangled also for red detunings $\Delta>-\omega_R$. Naively, one could expect that $\tilde{\delta}=-\Delta-\omega_R=0$ should generate maximum BEC-BEC entanglement since then the effective atom-atom coupling is the largest. However, as clear from Fig.~\ref{time_avg_no_aux} (b) the picture is more complex; a strong atom-field coupling implies also strong back-action of the field onto the atoms (and also an increased influence of the photon reservoir on the atom-atom coherence), which can counteract build-up of BEC-BEC entanglement. This tradeoff between increasing the effective atom-atom interaction and decreasing the influence of the cavity field and its decoherence is also reflected in the non-monotonous behavior of the negativity as a function of $g/g_c$. As shown in Fig.~\ref{time_avg_no_aux} (c) due to entanglement sharing  constraints in a tri-partite  quantum system, the BEC-BEC entanglement grows at the expense of the cavity-BEC entanglement.

\subsection{Inferring entanglement} 

\begin{figure}[h!]
  \centering
    \includegraphics[width=1\textwidth]{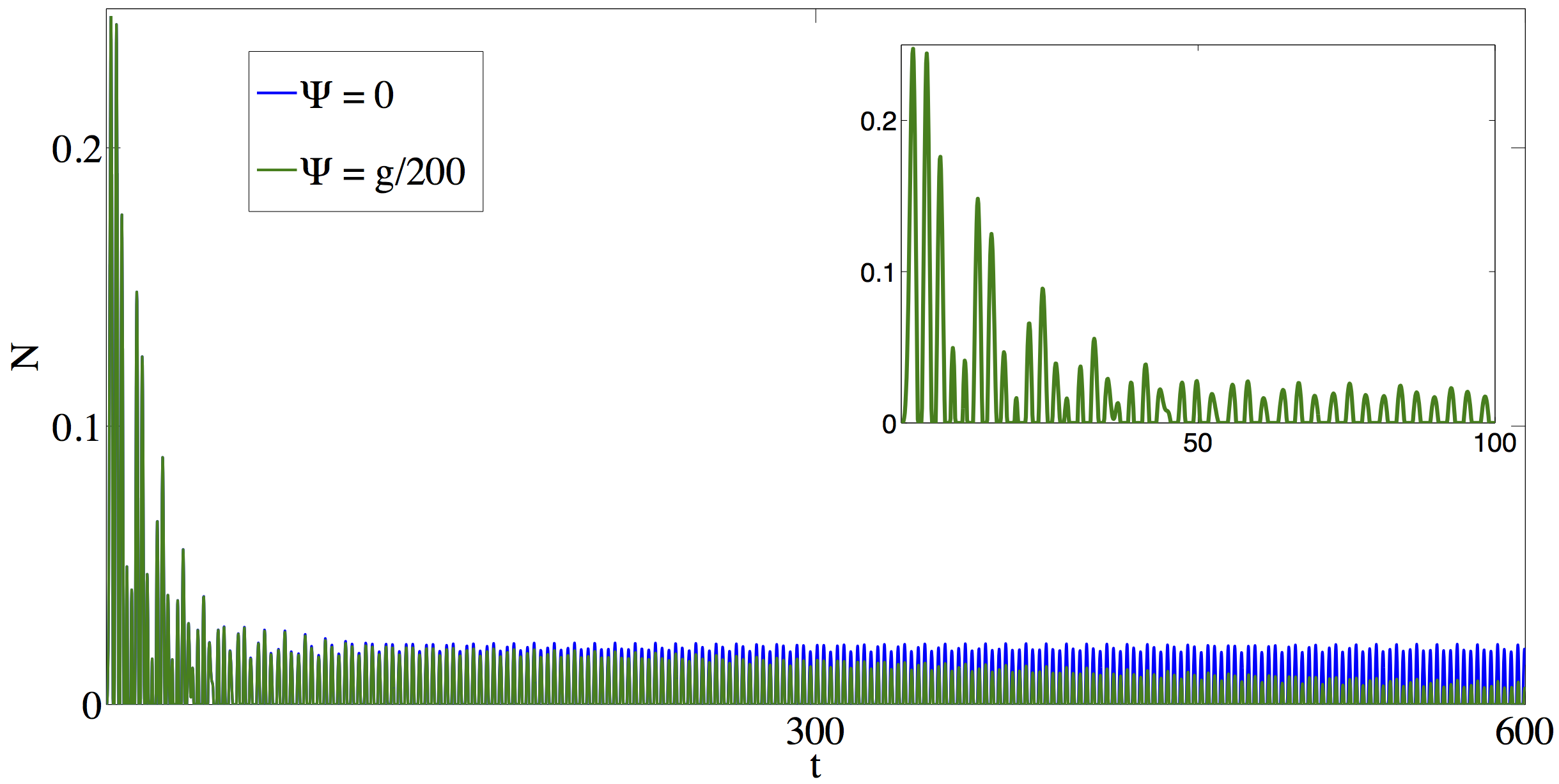}
   \caption{Time-evolution of the logarithmic negativity $\rm{N}$, serving as a measure of entanglement, between the atomic $\hat b$ and $\hat c$ modes in the presence, $\Psi\neq0$, (green) and absence, $\Psi=0$ (blue) of the auxiliary $\hat w$ mode. Other parameters include $\Delta=-1$, $\omega_{R}=1$, $\kappa=0.05$, $\gamma=0.05\kappa$, and $g/g_{c}=1.05$. In the presence of the auxiliary $\hat{w}$ mode the logarithmic negativity $\rm{N}$ monotonically decreases with time. The inset shows transient oscillations of $\rm{N}_{\Psi = 0}$ and $\rm{N}_{\Psi \neq 0}$ featuring multiple occurrences of   entanglement sudden death and sudden birth. At these short times, the influence of the decay of the $\hat w$ mode is not apparent; we note that the characteristic time scale for this decay $\gamma^{-1}=400$. Note that with $\gamma=\kappa=0.05$ we would still observe entanglement for $t\lesssim50$ and thereby also several deaths/births of the entanglement.}
   
 \label{time_evolution_disp_inference}
\end{figure}

A challenging aspect of any scheme involving macroscopic entanglement generation is the actual experimental detection of entanglement. In this section we will propose  a strategy to indirectly infer the degree of  entanglement generated  between the two atomic BECs. The scheme can be used to indirectly infer the degree of entanglement generated between the two BECs in the superradiant phase for small values of the critical coupling ratio $g/g_{c} \gtrsim 1$. More precisely, we suggest to make use of an additional (auxiliary) optical $\hat w$ mode to indirectly infer the degree of entanglement between the two atomic  samples. The auxiliary mode couples to the relative atomic  $\hat s$ mode with a strength $\Psi$ as
\begin{eqnarray}\label{auxham}
\tilde{H}_{\rm sr}^{\rm lin-aux}=\omega_{p}\hat{p}^{\dagger}\hat{p}+\omega_{q}\hat{q}^{\dagger}\hat{q}+\Omega\hat{s}^{\dagger}\hat{s}+\frac{(\Omega+\Delta)}{2}(\hat{p}^{\dagger}\hat{q}+\hat{q}^{\dagger}\hat{p})
+\zeta (\hat{p}\hat{q}+\hat{q}^{\dagger}\hat{p}^{\dagger})+\zeta(\hat{s}^{2}+\hat{s}^{\dagger 2})\nonumber \\
+\left(\frac{\zeta}{2}+\frac{\phi}{\sqrt{2}}\right)(\hat{p}^{2}+\hat{p}^{\dagger 2})+\left(\frac{\zeta}{2}-\frac{\phi}{\sqrt{2}}\right)(\hat{q}^{2}+\hat{q}^{\dagger 2})
+\Omega_{w}\hat{w}^{\dagger}\hat{w}+\Psi(\hat{s}^{\dagger}\hat{w}+\hat{w}^{\dagger}\hat{s}).
\end{eqnarray}
A discussion on an actual physical  realization of the above Hamiltonian has been left as an open question which might be studied in a future work. It should be pointed out that the Hamiltonian \eqref{auxham} has a {\it beam-splitter} (passive) interaction between the relative $\hat s$ mode and the auxiliary $\hat w$ mode~\cite{meystre}. In this sense, the $\hat w$ mode initially  prepared in a classical state and evolving under the Hamiltonian  \eqref{auxham}  can only become non-classical as a result of a quantum state swap with the $\hat s$ mode. The open dynamics of our hybrid tri-partite quantum system, now interacting with an auxiliary $\hat w$ mode, is described  by the following  master equation  
\begin{equation}
\frac{d}{dt}\hat\rho=-i[\hat{H}_{\rm sr}^{\rm lin-aux},\hat\rho]+\frac{\kappa}{2}\mathcal L_{\hat{p}}[\hat\rho]+\frac{\kappa}{2}\mathcal L_{\hat{q}}[\hat\rho]
-\frac{\kappa}{2}(2 \hat p \rho \hat q^\dagger - \hat q^\dagger \hat p \rho - \rho \hat q^\dagger \hat p)
-\frac{\kappa}{2}(2 \hat q \rho \hat p^\dagger - \hat p^\dagger \hat q \rho - \rho \hat p^\dagger \hat q)
 +\gamma\mathcal L_{\hat{w}}[\hat\rho].
\end{equation}
Here we have included photon decay of the auxiliary mode with a loss rate $\gamma$. It should be pointed out that a damped auxiliary $\hat w$ mode provides 
an indirect dissipation channel for the $\hat s$ mode. In other words, indirect damping introduced by the $\hat w$ mode can allow the $\hat s$ mode to settle to a time-independent steady state. Fig.~\ref{time_evolution_disp_inference}  shows  transient bi-partite entanglement between the atomic $\hat b$ and $\hat c$ modes in the presence and absence of the auxiliary $\hat w$ mode. The time decaying envelope of the quantum entanglement between the atomic modes can be attributed to the indirect damping suffered by the $\hat s$ mode. More precisely, without losses of the $\hat s$ mode the entanglement does not reach a vanishing steady state -- the sustainable quantum correlations derive from the lossless $\hat s$ mode. With losses of the $\hat w$ mode included, however, the BEC-BEC entanglement vanishes in the long time limit. The Fig.~\ref{time_evolution_disp_inference} also gives a clear signature  of entanglement sudden death and birth \cite{ting04}, arising from the common coupling of the two atomic BECs to a single cavity mode. As shown in Fig.~\ref{time_evolution_disp_inference}, entanglement sudden death and birth occur both in the absence and presence of the auxiliary $\hat w$ mode. Similar observation of entanglement sudden death and birth has been reported in a scheme proposing entanglement generation among distant optomechanical systems \cite{cjos12}. 
 
\begin{figure}[h!]
  \centering
    \includegraphics[width=1\textwidth]{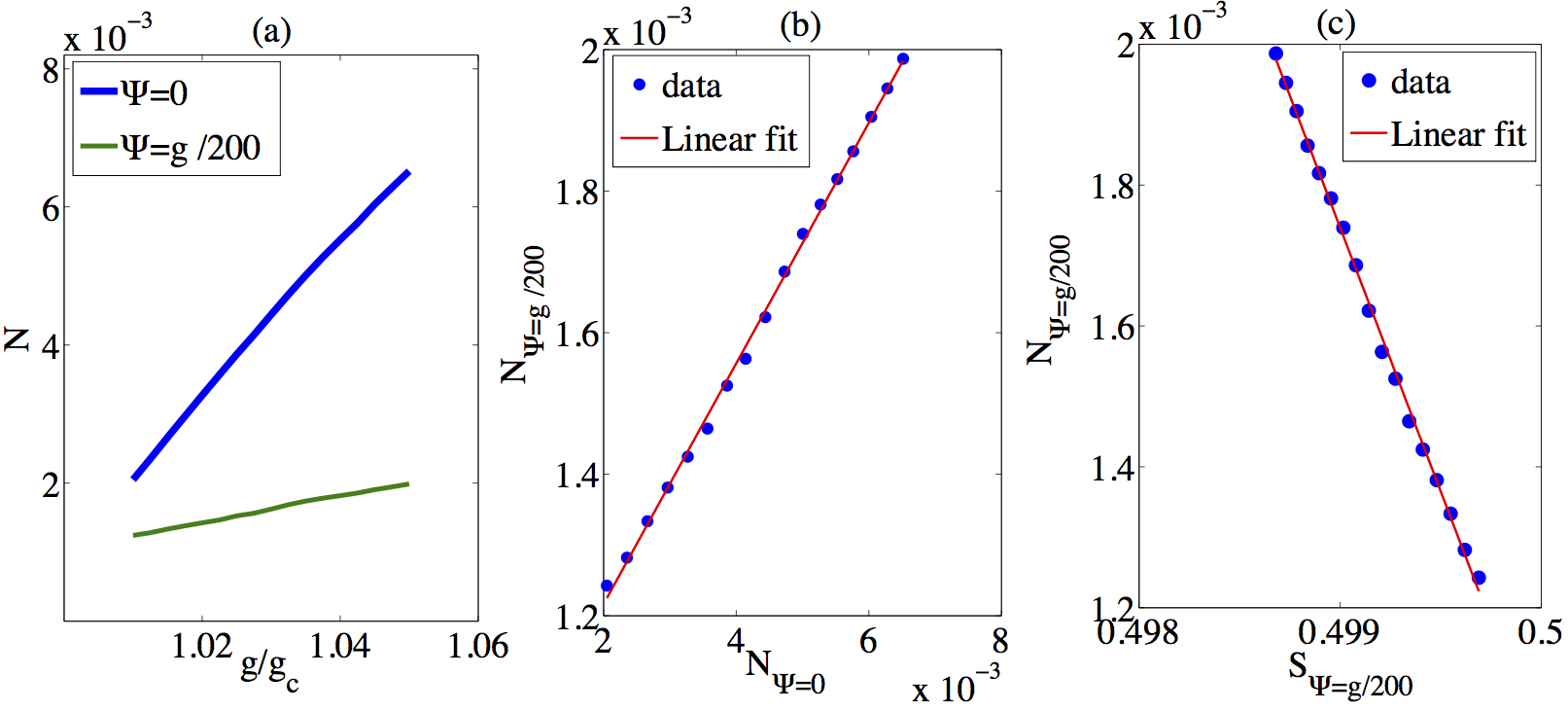}
   \caption{Entanglement (logarithmic negativity) between the two atomic $\hat b$ and $\hat c$ modes as a function of $g/g_c$ (a) in the presence ($\Psi\neq0$) and absence ($\Psi=0$) of the auxiliary $\hat w$ mode. The linear dependences in the two cases lead to proportionality shown in (b);  $\rm N_{\Psi\neq0}$ vs. $\rm N_{\Psi=0}$. Dots are numerical data, while the solid line a least square linear fit. In (c) the logarithmic negativity is displayed vs the single mode squeezing $S_{\Psi\neq0}$ of the auxiliary mode. As is clear, increasing the atom-atom entanglement implies an increased squeezing of the $\hat w$ mode, which suggests that squeezing would be an indirect measure of entanglement in the system. Other parameters include $\Delta=-1$, $\omega_{R}=1$, $\kappa=0.05$, and $\gamma=0.05\kappa$. 
   }
 \label{ent_disp_inference}
\end{figure}

In Fig.~\ref{ent_disp_inference} (a) we plot the logarithmic negativity both with ($\Psi\neq0$) and without ($\Psi=0$) the auxiliary mode vs. small values of the critical coupling ratio $g/g_{c} \gtrsim 1$. The linear dependence in this parameter regime suggests a simple one-to-one mapping between the two cases. Indeed, we find that for values of the critical ratio $g/g_{c}\gtrsim 1$, the  time-averaged quantum entanglement between the $\hat b$ and $\hat c$ modes in the absence of the auxiliary $\hat{w}$ mode ($\rm{N}_{\Psi = 0}$) is proportional to the same quantity in the presence of auxiliary $\hat{w}$ mode ($\rm{N}_{\Psi \neq 0}$). Thus, the effect of the auxiliary mode on the atom-atom entanglement is to just scale it. This feature is captured in Fig.~\ref{ent_disp_inference} (b). Moreover, the single mode squeezing $S_{\Psi \neq 0}$~\cite{mw} of the $\hat w$ mode is also found to behave linearly with the bi-partite entanglement between the $\hat b$ and $\hat c$ modes, {\it i.e.} $\rm{N}_{\Psi \neq 0}\propto S_{\Psi \neq 0}$ as shown in Fig.~\ref{ent_disp_inference} (c). The connection between entanglement and squeezing has long been known~\cite{sorensen}, and as the later can be determined by means of homodyne measurements on the auxiliary  $\hat w$ mode~\cite{mw}, this allows the possibility of indirectly quantifying the macroscopic entanglement between the two BECs. It should be possible to subject the degree of single mode squeezing to a post-processing procedure and one can then rescale the bi-partite entanglement ($\rm{N}_{\Psi = 0}$) according to the linear behavior shown in Fig.~\ref{ent_disp_inference} (b) and (c).  
 
\section{Conclusions}

In this work we have explored a possibility to generate macroscopic entanglement between two BEC atomic samples. We considered a physical setting where two laser driven spatially separated BECs are placed in a high finesse optical cavity where the photons can leak out of the cavity at a rate $\kappa$. The two BEC samples interact indirectly through the same cavity field imposing an indirect atom-atom interaction which generates the entanglement. In the thermodynamic limit our physical model is known to exhibit a Dicke-type phase transition when the light-matter coupling exceeds a critical value. This phase transition separates the normal phase from a superradiant phase. Starting with a semiclassical analysis of the steady state of our driven-dissipative model, and adding a linearized treatment of quantum fluctuations using the Holstein-Primakoff representation we solve the resulting master equations in the two phases separately. We find that the two BEC samples remain separable in the normal phase but can become entangled in the superradiant phase. This generation of bi-partite entanglement between the two BECs can be seen as a signature  of our hybrid system entering  the superradiant phase where  quantum coherence persists over macroscopic length scales. In particular, by entering the superradiant phase the cavity-BEC entanglement peaks at the critical point and almost no BEC-BEC entanglement exists at this point. Deeper in the superradiant phase the phenomenon of entanglement sharing occurs where the cavity-BEC entanglement is lowered while the BEC-BEC entanglement gets stronger. Even though the open-driven system does not approach a steady state in the general case, the time-averaged BEC-BEC entanglement is persistent. Since the two BECs are coupled to the same leaky photon mode, we also encounter an entanglement sudden death/birth evolution. We have also outlined a strategy to indirectly infer the degree of entanglement between the two atomic BECs. This scheme is no direct measure of the entanglement between the two condensates, but it should be a clear smoking gun of it.   

\acknowledgments{Acknowledgments}
Tobias Donner is thanked for helpful discussions on the experimental realization part. C.J. acknowledges the York Centre for Quantum Technologies (YCQT) Fellowship for financial support, while J. L. acknowledges VR-Vetenskapsr\aa det (The Swedish Research Council) and KAW (The Knut and Alice Wallenberg foundation) for financial support.


\authorcontributions{Author Contributions}
Jonas Larson conceived the main idea of the project, while Chaitanya Joshi carried out most of the theoretical and numerical calculations. Both authors contributed to writing and revising the manuscript.


\conflictofinterests{Conflicts of Interest}

The authors declare no conflict of interest.

\bibliographystyle{mdpi}
\makeatletter
\renewcommand\@biblabel[1]{#1. }
\makeatother



%


%

\end{document}